\documentclass[a4paper,12pt]{article}
\usepackage{epsfig,float}
\usepackage[left=2cm,right=2cm]{geometry}
\usepackage{a4wide,graphicx}
\usepackage[sort&compress,square,comma,numbers]{natbib}

\newcommand{\Ima}{\textrm{Im}}

\newcommand{\mev}{\textrm{ MeV}}
\newcommand{\be}{\begin{equation}}
\newcommand{\ee}{\end{equation}}
\newcommand{\ba}{\begin{eqnarray}}
\newcommand{\ea}{\end{eqnarray}}
\newcommand{\gev}{\textrm{ GeV}}

\begin{document}

\title{$J/\psi$ reaction mechanisms and suppression in the nuclear medium}

\author{R. Molina, C. W. Xiao and E. Oset\\
{\small{\it $^1$Departamento de F\'{\i}sica Te\'orica and IFIC,
Centro Mixto Universidad de Valencia-CSIC,}}\\
{\small{\it Institutos de
Investigaci\'on de Paterna, Apartado 22085, 46071 Valencia, Spain}}\\
\\}

\date{}

\maketitle
\begin{abstract}
Recent studies of the interaction of vector mesons with nuclei make possible and opportune the study of the interaction of the $J/\psi$ with nuclei and the investigation of the origin of the $J/\psi$ suppression in its propagation thorough a nuclear medium. We observe that the transition of  $J/\psi N$ to $VN$ with $V$ being a light vector, $\rho, \omega,\phi$, together with the inelastic channels, 
$J/\psi N \to \bar D \Lambda_c$ and $J/\psi N \to \bar D \Sigma_c$ leads to a particular shape of the inelastic cross section.
Analogously, we consider the mechanisms where the exchanged $D$ collides with a nucleon and gives $\pi \Lambda_c$ or $\pi \Sigma_c$. The cross section has a peak around $\sqrt s=4415$ MeV, where the $J/\psi N$ couples to a resonance predicted recently. We study the transparency ratio for electron induced $J/\psi$ production in nuclei at about 10 GeV and find that 30 - 35 \% of the $J/\psi$ produced in heavy nuclei are absorbed inside the nucleus. This ratio is in line with depletions of $J/\psi$ though matter observed in other reactions.

\end{abstract}

\section{Introduction}
\label{Intro}

The subject of $J/\psi$ suppression in nuclei has a long history \cite{Vogt:1999cu} and many plausible reasons for it have been given. Reaction mechanisms of $J/\psi$ with the nucleons are suggested in \cite{Vogt:2001ky,Kopeliovich:1991pu,Sibirtsev:2000aw}. Parton shadowing in the target nucleus may suppress the probability of
producing a $J/\psi$ \cite{Eskola:2009uj}. Energy loss of the incident parton in the nuclear medium, prior to $c \bar c$ production, may  alter the $J/\psi$ production cross section \cite{Gavin:1991qk,Marco:1997xb}. Also, a suppression of the $J/\psi$ has been proposed as a signature of the formation of Quark-Gluon
Plasma in ultrarelativistic nucleus-nucleus collisions \cite{Matsui:1986dk}. The reaction mechanisms producing the $J/\psi$ in a first place are also not well understood \cite{Brambilla:2010cs}. In any case, a proper understanding of what happens in hot nuclear matter in  
ultrarelativistic nucleus-nucleus collisions demands that we understand what happens and why in cold matter as mentioned in \cite{Alessandro:2006jt}. In this sense $J/\psi$ suppression has been extensively searched in p-nucleus collisions in several fixed target experiments (NA3 \cite{Badier:1983dg}, E772 \cite{Alde:1990wa}, NA38 \cite{Abreu:1998ee}, E866 \cite{Leitch:1999ea},
E672/E706 \cite{Gribushin:1999ha}, NA50 \cite{Alessandro:2003pi,Alessandro:2003pc,Alessandro:2006jt} and more recently in NA60 \cite{Arnaldi:2010ky}.

  Our aim in this work is to exploit recent progress in the theoretical description of the interaction of vectors mesons with nucleons and apply these ideas to study mechanisms of $J/\psi$ absorption in nuclei. We have in mind the depletion of $J/\psi$ in production reactions in nuclei induced by elementary particles, protons, photons, etc. The starting point is to recall recent advances on our theoretical understanding of the interaction of vector mesons with nucleons. At small and intermediate energies the practical tool to deal with vector meson interactions is the use of effective Lagrangians of the local hidden gauge theory  \cite{hidden1,hidden2,hidden3,hidden4} which incorporate pseudoscalar mesons, vector mesons and photons. Concerning the pseudoscalar interaction these Lagrangians are equivalent to the chiral Lagrangians 
\cite{Weinberg:1968de,Gasser:1983yg} assuming vector meson dominance, thus, they account for chiral symmetry. In addition they allow to extend the theory to provide the interaction of pseudoscalar mesons with vector mesons and vector mesons with themselves. If one considers the coupling of vector mesons to baryons \cite{Klingl:1997kf,Palomar:2002hk} one can then address the interaction of vectors with baryons. Yet, even at low energies the use of perturbation theory becomes inadequate and nonperturbative techniques are demanded to study this interaction. By combining the information from the Lagrangians and unitary in coupled channels, following the pattern of the chiral unitary approach  \cite{Oller:2000ma}, a study of the vector-baryon interaction is done in \cite{angelsvec} for the case of the baryons of the octet of the proton and in \cite{souravvec} for the case of the baryons of the decuplet of the $\Delta$. It is found there that several resonances appear as a consequence of the interaction which can be associated to known states of the PDG \cite{pdg}. The extrapolation of these works to the charm sector was done in \cite{Wu:2010jy,Wu:2010vk}, where some $N^{*}$ and $\Lambda^*$ resonances in the hidden charm sector were dynamically generated from $DN$ and other coupled channels, $\pi\Sigma_c$ and $\pi\Lambda_c$ among them.
This works contain the tools to address the $J/\psi N$ interaction which are used here. 

    Furthermore, when it comes to study the propagation of vector mesons with nuclei we apply also recent tools developed in the study of the $\bar K^*(890)$ in nuclei \cite{laura}. This latter work has gone one step forward with respect to the well established works on the issue \cite{Rapp:1997fs,Peters:1997va,Rapp:1999ej,Urban:1999im,Cabrera:2000dx,Cabrera:2002hc} that were constructed to address the problem of vector meson propagation through nuclei \cite{Hayano:2008vn,Leupold:2009kz}. While the latter quoted works concentrated mostly on the modification of the decay channels and the coupling to some resonance-hole components introduced empirically, the dynamical generation of these resonances, to which the vector-nucleon couples so strongly, in the work of \cite{angelsvec}, allows to address the problem from a more microscopical point of view. Indeed, in \cite{laura} there are two sources of vector modification of the $\bar K^*$, the modification of the particles of its decay channel, $\pi \bar K$, and the $\bar K~N$ interaction modified in the medium, which is studied nonperturbatively in \cite{laura} and gives rise to dynamically generated resonances in the region of 2000 MeV. In this sense the coupling of the $\bar K^*$ to hole-resonance components is done automatically, with the strength provided by the same model. This of course has more relevance when we go to the charm sector since experimental information on baryonic resonances is scarce and their coupling to vector meson components is not known.

In the case of the $\rho$ meson the decay channel is $\pi \pi$, and $\pi \bar K$ for the $\bar K^*$. The equivalent mesonic decay channel of the $J/\psi$ is $D \bar D$, but it is closed kinematically. Yet, in the medium there is more available energy for the opening of new decay channels. Indeed, the channel $\bar{D}\Lambda_c$ is slightly above to the $J/\psi N$ threshold and can lead to absorption phenomena in the medium. The extrapolation to SU(4) of the coupling of vector mesons to pseudoscalars, as done in \cite{Molina:2009ct}, provides a strong coupling of 
$J/\psi$ to $D \bar D$, and the medium related decay channels, with $D N \to \Lambda_c$ or $D N \to \Sigma_c$,  are studied in the present work. When implementing vertex corrections in the medium, a contact term $J/\psi N\to\bar{D}\Lambda_c$, which is called Kroll Rudermann term, must also be taken into account. Altogheter, this leads to a relevant source of the $ J/\psi$ absorption in the medium through the reaction $J/\psi N\to \bar{D}\Lambda_c$. In addition, one can also consider the creation of one pion in the final state, i. e. $J/\psi N\to \bar{D}\Sigma_c\pi,\bar{D}\Lambda_c\pi$. This reaction requires more energy, however, it is interesting to study it since the $\pi\Sigma_c,\pi\Lambda_c$ channels are decay channels of the $\Lambda_c(2595)$ and $\Sigma_c(2800)$ resonances respectively, which are dynamically generated \cite{Hofmann:2005sw,Mizutani:2006vq,Tolos:2007vh}. 

  In order to test the relevance of the $J/\psi$ absorption mechanisms found, we evaluate the transparency ratio for photoproduction of $J/\psi$ in nuclei. Using beams of around 10 GeV, and energy accessible in the Jefferson Lab upgrade, we look for the rate of production in different nuclei and we find a depletion of about 30 - 35 \% for heavy nuclei. Although we do not want to venture into very high energies, we note however that this is the order of magnitude for the suppression found at higher energies in $p$-nucleus collisions.

\section{Formalism}

\subsection{Vector-baryon coupled channels approach}
\label{vbcca}

Recently, a study of the vector-baryon interaction in the hidden charm sector around energies of $4$ GeV, has been tackled in \cite{Wu:2010jy,Wu:2010vk}. In the sector with isospin $I=1/2$ and strangeness $S=0$, three channels are considered: $\bar{D}^*\Lambda_c$, $\bar{D}^*\Sigma_c$ and $J/\psi N$. The potential is evaluated using an SU(4) extrapolation of the local hidden gauge approach with symmetry breaking ingredients implemented \cite{Wu:2010jy,Wu:2010vk}. The amplitudes of Feynmann diagrams like those in Fig. \ref{fig:diag} a) are evaluated, and the potential after projecting in s-wave takes the form:
\begin{figure}[htpb]
\begin{center}
\includegraphics[scale=0.6]{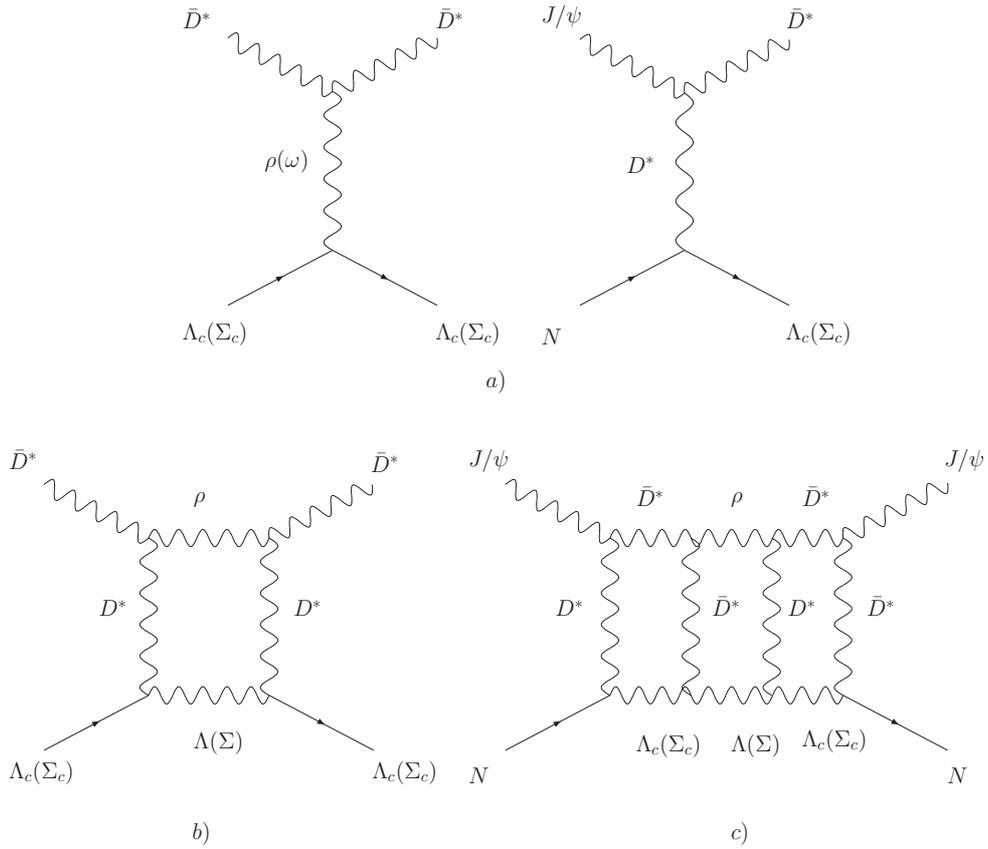}
\end{center}
\caption{a) Vector exchange diagrams for the vector-baryon interaction considered in \cite{Wu:2010jy,Wu:2010vk}. b) Box diagram with $\rho\Lambda(\Sigma)$ in the intermediate state. c) $J/\psi N\to J/\psi N$ like-box diagram with $\rho \Lambda(\Sigma)$ in the intermediate state.}\label{fig:diag}
\end{figure}

\be
V_{ij}^{WT}=C_{ij} \frac{1}{4f^2} (E + E') \vec{\epsilon}\, \vec{\epsilon}\,',\label{vij}
\ee
for $i,j=\bar{D}^*\Lambda_c,\bar{D}^*\Sigma_c$. In the above formula, $E$, $E'$ are the energies of vector mesons, $\vec{\epsilon}$,  $\vec{\epsilon}\,'$ the polarization vectors, $f\equiv f_\pi=93\mev$, and $C_{ij}$ are coefficients given in \cite{Wu:2010jy,Wu:2010vk}. The transition between these two channels is achieved through the exchange of one $\rho$ or $\omega$ mesons. For transitions between $\bar{D}^*\Lambda_c,\bar{D}^*\Sigma_c$ and $J/\psi N$, the full propagator of the $D^*$ meson is taken into account. Thus, we have
\be
V_{kl}^{WT}(J/\psi N\to \bar{D}^*\Lambda_c,\bar{D}^*\Sigma_c)=-\frac{C_{kl} g^2}{p^2_{D^*}-m^2_{D^*}}(E_{D^*}+E_{J/\psi})\vec{\epsilon}\, \vec{\epsilon}\,',\label{vkl}
\ee
where $g=m_\rho/2f$. Note that the vertices $J/\psi J/\psi \omega$ or $J/\psi J/\psi \rho$ are forbidden for G-parity and isospin respectively, which leads to a zero potential of the tree order amplitude $J/\psi N\to J/\psi N$. But, when amplitudes are unitarized via the Bethe Salpeter equation, the resumation of loops implies indirect reactions $J\psi N\to \bar{D}^*\Lambda_c(\Sigma_c)\to J\psi N$.

The scattering matrix is given by the Bethe-Salpeter equation in coupled channels,
\be 
T=[1-\tilde{V} G]^{-1} \tilde{V}  \vec{\epsilon}\, \vec{\epsilon}\,',\label{tmatrix}
\ee
where $G$ is the diagonal matrix for the loop function of intermediate VB propagators given in \cite{Wu:2010jy,Wu:2010vk} and $\tilde{V}$ is the potential of Eq. \ref{vij} removing $\vec{\epsilon}\, \vec{\epsilon}\,'$. When going to the complex plane of the energy, one resonance is found at the position $\mathrm{Re}(\sqrt{s})=4415$ MeV. Pole positions and couplings to the different channels are given in Table \ref{tab:polepos}.
\begin{table}[htpb]\setlength{\tabcolsep}{0.20cm} \renewcommand{\arraystretch}{1.5}             
 \begin{center}
  \begin{tabular}{llrrr}
   \hline\hline
$(I,S)$&$\sqrt{s}=4415-9.5i$&\multicolumn{3}{r}{Channels}\\
\hline
$(1/2,0)$& &$\bar{D}^*\Sigma_c$&$\bar{D}^*\Lambda_c$&$J/\psi N$\\\hline
 $g_a$& &$2.83-0.19i$&$-0.07+0.05i$&$-0.85+0.02i$\\
\hline\hline
  \end{tabular}

 \end{center}
\caption{Pole position and coupling constants ($g_a$) to various channels for the state found in the sector $(I,S)=(1/2,0)$}\label{tab:polepos}
\end{table}
In addition, there can be transitions from the heavy vector-heavy baryon channels to light vector-light baryon channels   with a big momentum transfer to the last ones for the energies that we consider. To account for this momentum dependence, the light vector-light baryon channels are implemented through box Feynmann diagrams, see Fig. \ref{fig:diag} b). This is done because the masses of the intermediate channels are very far from the energies under consideration for $J/\psi N$. This transition potential is derived from the same hidden gauge Lagrangians, and it is given by \cite{Wu:2010jy,Wu:2010vk}

\be 
\delta \tilde{V}_{ab}^{Box} = \sum \limits_c \tilde{V}_{al} \, G_l \, \tilde{V}_{lb},
\ee
where $l$ stands for the light channels $\rho N$, $\omega N$, $\phi N$, $K^* \Lambda$, $K^* \Sigma$, and
\begin{eqnarray}
\tilde{V}_{al}&= &-C_{al} g^2 \frac{-2E_{V_{1}}+\frac{(M_{B_{3}}-M_{B_{1}})(M^2_{V_{1}}+M^2_{V^*_{1}}-M^2_{V_{3}})}{M^{2}_{V^{*}_{1}}}}
{M^2_{V_{1}}+M^2_{V_{3}}-2E_{V_{3}}E_{V_{1}}-M^2_{V^*_{1}}}\label{eq:box}
\end{eqnarray}
 Here $l$ stands for a different group of
$V_{3}B_{3}$, and $C_{al}$ given in the Table \ref{cij} of the Appendix. Then, the kernel $V$ in the Bethe Salpeter
equation, Eq. (\ref{tmatrix}), becomes now:
\begin{eqnarray}
V_{ab}(V_{1}B_{1}\rightarrow
V_{2}B_{2})&=&V_{ab}^{WT}+\delta \tilde{V}_{ab}^{Box}\label{widpb2}
\end{eqnarray}
with $V_{ab}^{WT}$ given by Eqs. (\ref{vij}) and (\ref{vkl}). Since the light vector-light baryon intermediate channels are very far from the thresholds of $J/\psi N, \bar{D}^*\Lambda_c(\Sigma_c)$, the real part of the box diagrams is small and only the imaginary part matters, but one pays the prize of having the factor $-m^2_{D^*}$  in the denominator of the propagator, which reduces its contribution. Thus, the effect of the inclusion of the potential $\delta \tilde{V}_{acb}^{Box} $ in the Bethe Salpeter equation is only a moderate widening of the resonance. With this, the state found with mass $M=4415$ MeV has a width of $28$ MeV added to the $19.2$ MeV due to its decay into the $J/\psi N$ channel, which results in a total width of around $50$ MeV \cite{Wu:2010jy,Wu:2010vk}. The fact that this width is small for a state with such high mass is due to the fact that the transitions are mediated by a heavy vector meson. It is worth noting that the $J/\psi N$ channel, which concerns us in the present article, only can go to the light vector-light baryon channels through intermediate states with $\bar{D}^*\Lambda_c$, $\bar{D}^*\Sigma_c$, see Fig. \ref{fig:diag} c).
Since the depletion has to do with the inelastic $J/\psi N$ cross section, we evaluate it by using the optical theorem that states in our normalization
\be 
\sigma_{tot} = -\frac{M_N}{P_{CM}^{J/\psi} \sqrt{s}} \Ima\;T_{J/\psi N \to J/\psi N},\label{eq:sigtot}
\ee
hence, by evaluating also the elastic cross section we have
\begin{eqnarray}
\sigma_{in} &&= \sigma_{tot} - \sigma_{el} \\
&&=-\frac{M_N}{P_{CM}^{J/\psi} \sqrt{s}} \Ima\;T_{J/\psi N \to J/\psi N} - \frac{1}{4 \pi} \frac{M_N^2}{s} \overline{\sum} \sum |T_{J/\psi N \to J/\psi N}|^2,\label{eq:sigin}
\end{eqnarray}
where $\sum, \, \overline{\sum}$ stand for sum and average over the spins of the nucleons and $J/\psi$.

\begin{figure}
\centering
\includegraphics[scale=0.8]{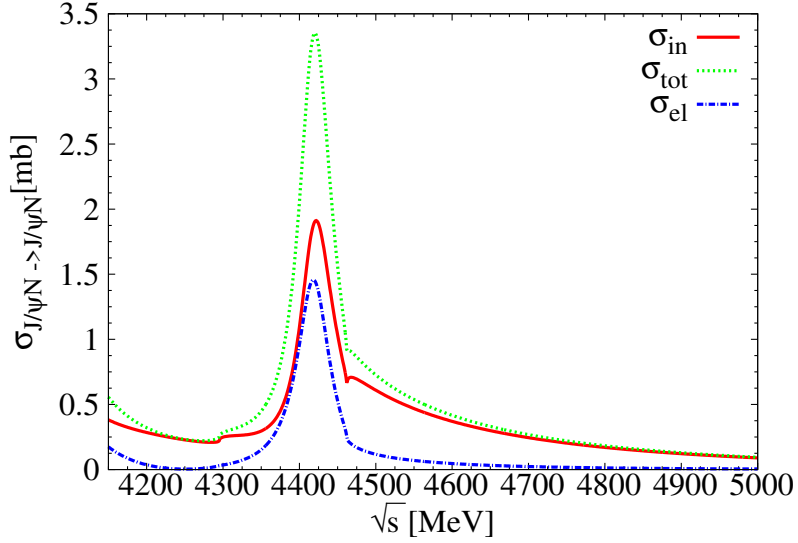}
\caption{The total, elastic and inelastic cross sections in Eqs. (\ref{eq:sigtot}) and (\ref{eq:sigin}).}\label{crosec}
\end{figure}

In Fig. \ref{crosec} we plot the results for these cross sections.
We observe a peak around $4425$ MeV, which corresponds to a hidden charm resonance found in \cite{Wu:2010jy,Wu:2010vk}. Actually, we are interested in the region of $J/\psi$ created in electron nucleus collisions for electrons around $10\gev$ which corresponds to $J/\psi$ moving in the rest frame of the nucleons with $\sqrt{s} \simeq 4050 - 5300 \mev$, which includes the resonant peak.

\subsection{The $J/\psi N \to \bar{D} \Lambda_c (\Sigma_c)$ reaction}
\label{JNDbLSc}

By analogy to the $\rho \to \pi\pi$ decay or $\bar{K}^*\to \bar{K}\pi$, the $J/\psi$ couples to $D\bar{D}$. Although the channel is not open for decay, the channels $J/\psi N\to \bar{D}\Lambda_c$, $\bar{D}\Sigma_c$ are nearly opened, the thresholds are  $4160$ and $4290$ MeV respectively, which requires a momentum $p^{\mathrm{cm}}_{J/\psi}=405$ MeV for $\bar{D}\Lambda_c$ production. The coupling $J/\psi D\bar{D}$ needed in these diagrams is obtained from the Lagrangian,
\be 
\mathcal{L}_{VPP} = -ig <[P,\partial_\mu P] V^\mu >,
\ee
with $g= \frac{M_V}{2f_\pi}$. For P and V we take the SU(4) matrices from \cite{gamermann},
\be 
P=\left(
\begin{array}{cccc}
\frac{\pi^0}{\sqrt{2}}+\frac{\eta}{\sqrt{6}}+\frac{\eta_c}{\sqrt{12}} & \pi^+ & K^+ & \bar{D}^0 \\
\pi^- & -\frac{\pi^0}{\sqrt{2}}+\frac{\eta}{\sqrt{6}}+\frac{\eta_c}{\sqrt{12}} & K^0 & D^- \\
K^- & \bar{K}^0 & \frac{-2\eta}{\sqrt{6}}+\frac{\eta_c}{\sqrt{12}} & D_s^- \\
D^0 & D^+ & D_s^+ & -\frac{3\eta_c}{\sqrt{12}}
\end{array}
\right),
\ee

\be 
V_\mu=\left(
\begin{array}{cccc}
\frac{\rho^0}{\sqrt{2}}+\frac{\omega}{\sqrt{6}}+\frac{J/\psi}{\sqrt{12}} & \rho^+ & K^{*+} & \bar{D}^{*0} \\
\rho^- & -\frac{\rho^0}{\sqrt{2}}+\frac{\omega}{\sqrt{6}}+\frac{J/\psi}{\sqrt{12}} & K^{*0} & D^{*-} \\
K^{*-} & \bar{K}^{*0} & \frac{-2\omega}{\sqrt{6}}+\frac{J/\psi}{\sqrt{12}} & D_s^{*-} \\
D^{*0} & D^{*+} & D_s^{*+} & -\frac{3J/\psi}{\sqrt{12}}
\end{array}
\right)_\mu.
\ee
The isospin doublets of $D$ are $(D^+,\,D^0), \, (-\bar{D}^0, \, D^-)$, and, thus, we find
\begin{eqnarray}
-i t_{J/\psi D^+(q) D^-(P-q)} &&=-i2gq_\mu \epsilon^\mu, \\
-i t_{J/\psi D^0(q) \bar{D}^0q} &&=-i2gq_\mu \epsilon^\mu, \\
-i t_{J/\psi D \bar{D}(I=0)} &&=-i2\sqrt{2}q_\mu \epsilon^\mu,
\end{eqnarray}
with $P$ the $J/\psi$ momentum.

\begin{figure}
\centering
\includegraphics[scale=0.6]{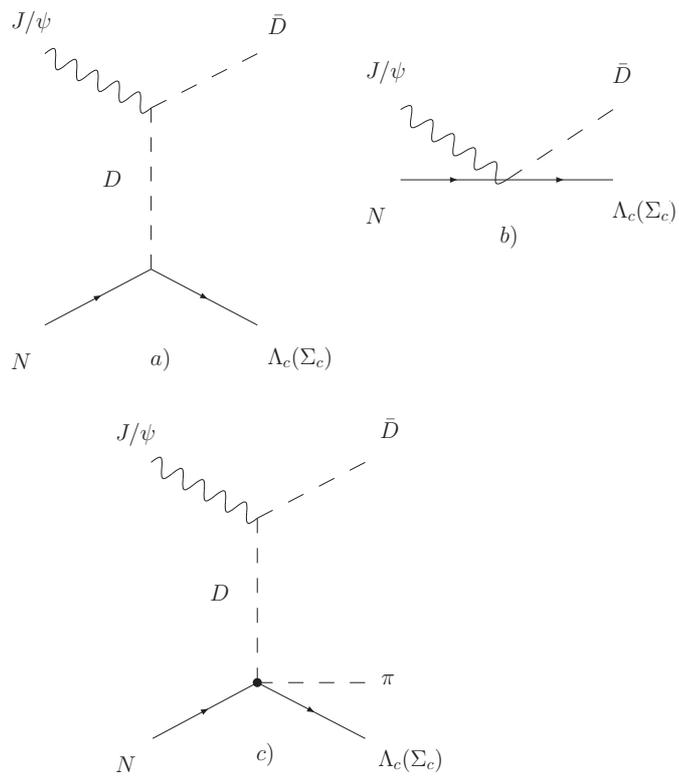}
\caption{Feynman diagrams of $J/\psi N \to \bar{D} \Lambda_c (\Sigma_c)$, a) Vector exchange contribution. b) Kroll Ruderman term. c) The $J/\psi N \to \bar{D} \pi\Lambda_c (\Sigma_c)$ reaction.}\label{figFeyn}
\end{figure}

We then evaluate the cross section for the Feynman diagrams of Figs. \ref{figFeyn} a).
This requires in addition the extension of the Yukawa vertex $DN \Lambda_c(\Sigma_c)$. One can use SU(4) symmetry or simply assume that the $D$ plays the analogous role as a $\bar{K}$ and $\Lambda_c (\Sigma_c)$ the role of $\Lambda (\Sigma)$. Then we find
\begin{eqnarray}
-i t_{D^0p \to \Lambda_c^+} &&= -\frac{1}{\sqrt{3}} \left( \frac{D+3F}{2f} \right) \vec{\sigma} \vec{q}, \\
-i t_{D^0p \to \Sigma_c^+} &&= \frac{D-F}{2f} \vec{\sigma} \vec{q}, \\
-i t_{D^+p \to \Sigma_c^{++}}& &= \sqrt{2} \frac{D-F}{2f} \vec{\sigma} \vec{q}.
\end{eqnarray}
We use the values $D=0.795,\; F=0.465$ \cite{Borasoy:1998pe}. The cross section for the process $J/\psi \to \bar{D}^0 \Lambda_c^+$ is given by
\be 
\sigma=\frac{M_N M_{\Lambda_c}}{4\pi}\frac{1}{s}\frac{p'}{p} \overline{\sum} \sum |T|^2,
\ee
where $p',\,p$ are the $\Lambda_c$ and $N$ momentum in the $J/\psi N$ CM frame and $|T|^2$ is given by
\begin{eqnarray}
\overline{\sum} \sum |T|^2 &&=  \frac{4}{3} g^2_D \Big[ \frac{(P\cdot p_{\bar{D}})^2}{M_{J/\psi}^2} - m_{\bar{D}}^2 \Big] \times \frac{1}{2} \frac{1}{m_N m_{\Lambda_c}} (m_N + m_{\Lambda_c})^2 \\
&& \times (pp' - m_N m_{\Lambda_c}) \times \frac{1}{(q^2 - m_D^2)^2} \times \frac{1}{3} \Big( \frac{3F+D}{2f} \Big)^2.
\end{eqnarray}
with $m_N$ the proton mass and $P,\;p_{\bar{D}}$ the four-momentum of the $J/\psi$ and $\bar{D}$ respectively and $g_D=m_{D^*}/2f_D$ ($f_D=206/\sqrt{2}$ MeV).

\begin{figure}
\centering
\includegraphics[scale=0.6]{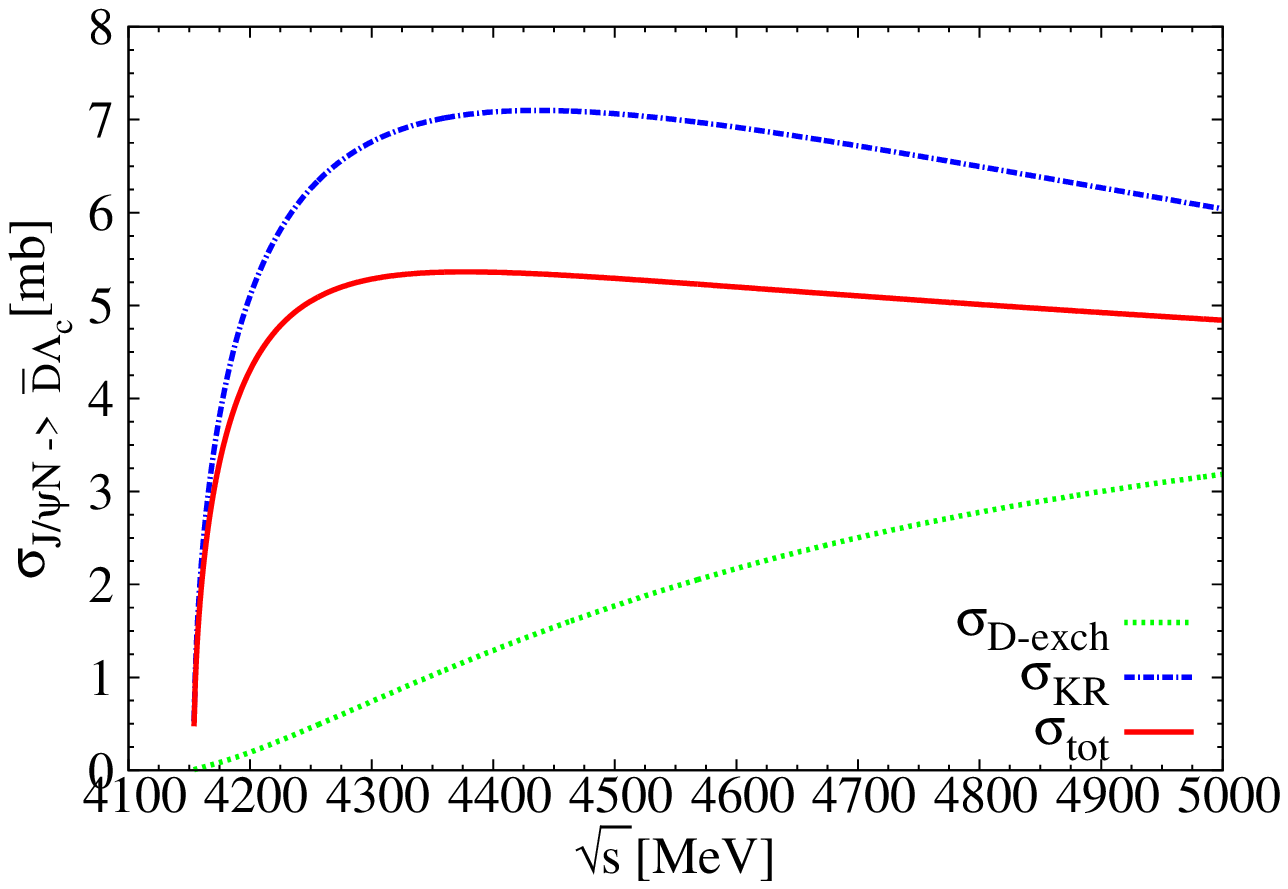}
\includegraphics[scale=0.6]{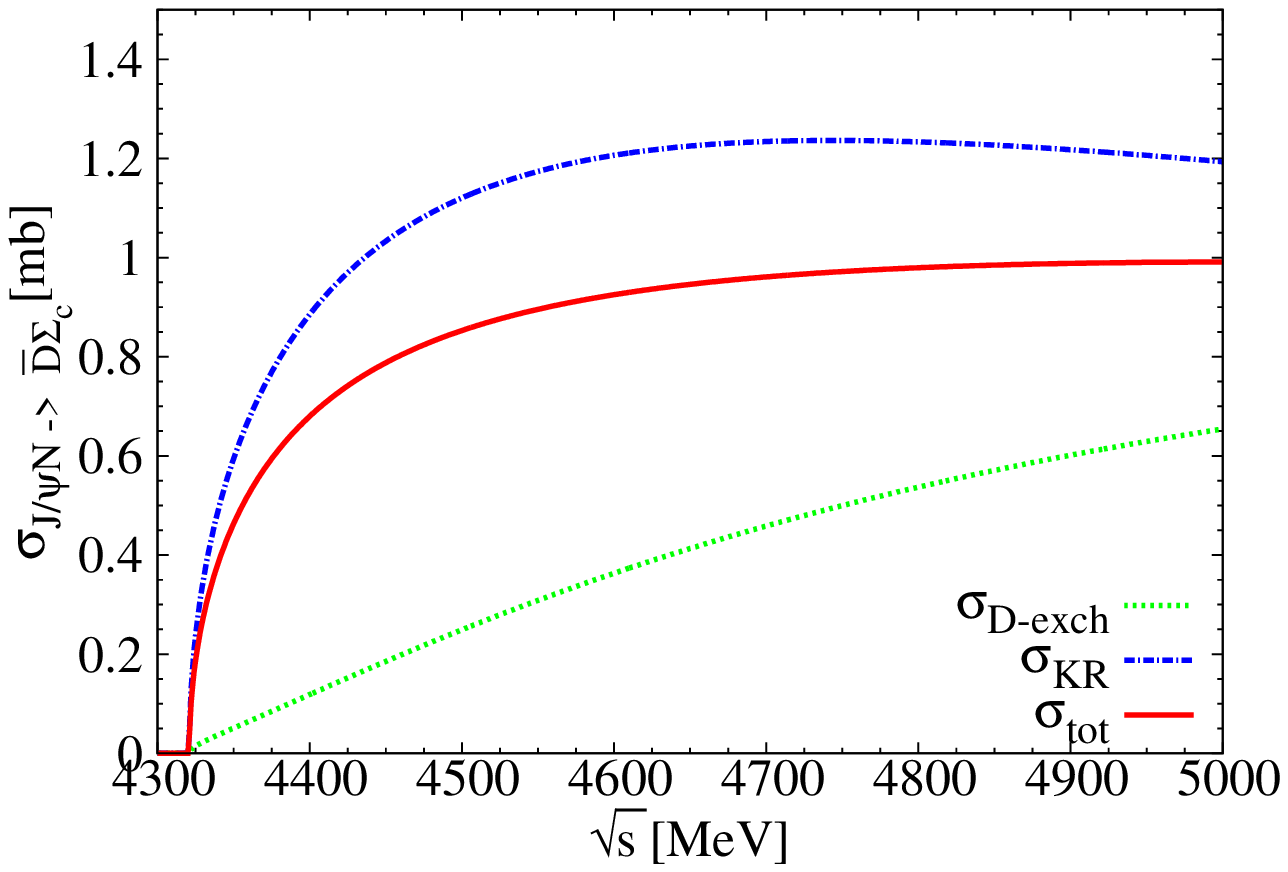}
\caption{The cross section for $J/\psi N \to \bar{D} \Lambda_c$ (left) and $J/\psi N\to \bar{D} \Sigma_c$ (right).}\label{figcro}
\end{figure}

For reasons of gauge invariance \cite{wambach,danirho,kanchan1,kanchan2,javier} one should add the Kroll Ruderman term, this is a contact term for the vector-two-baryon-pseudoscalar particles, see Fig. \ref{figFeyn} b). The prescription to get the Kroll Ruderman term is to substitute the meson pole term: $\vec{\epsilon} (\vec{P}_V + 2 \vec{q}) \frac{1}{(P_V+q)^2 - m_D^2} \vec{\sigma}(\vec{P}_V+q)$ by the Kroll Ruderman term: $\vec{\sigma} \vec{\epsilon}$. In the case of $J/\psi p\to \bar{D}^0\Lambda_c^+$ we get,
\be
-it_{p\Lambda_c^+J/\psi\bar{D}^0}=-\frac{g}{\sqrt{3}}\left( \frac{D+3F}{2f}\right) \vec{\sigma}\cdot\vec{\epsilon}
\ee
In Fig. \ref{figcro} we can see both contributions: $D$-exchange (dashed line), Kroll Ruderman (dot-dashed line), and the sum, which takes into interference (continuous line). Whereas the KR contribution remains constant while increasing $\sqrt{s}$, the $D$-exchange term increases with the momenta of the $J/\psi$. We observe that for energies around $4400$ MeV the KR term dominates, being about five times bigger than the $D$-exchange contribution, the latter has $\sigma\sim 1.2 $ mb around this energy for $\bar{D}\Lambda_c$. In the case of $\bar{D}\Sigma_c$, the sum is about one order of magnitude smaller than for $\bar{D}\Lambda_c$. The cross section for $J/\psi N \to \bar{D} \Lambda_c$ was also studied in \cite{Sibirtsev:2000aw} based on the same mechanism of Fig. \ref{figFeyn} a) and with similar results. We have also included here the Kroll Ruderman term following the developments of \cite{wambach,danirho,kanchan1,kanchan2,javier}.

\subsection{The $J/\psi N \to \bar{D}\pi\Lambda_c, \bar{D}\pi\Sigma_c$}
\label{JNDpLS}

\begin{figure}
\centering
\includegraphics[scale=0.6]{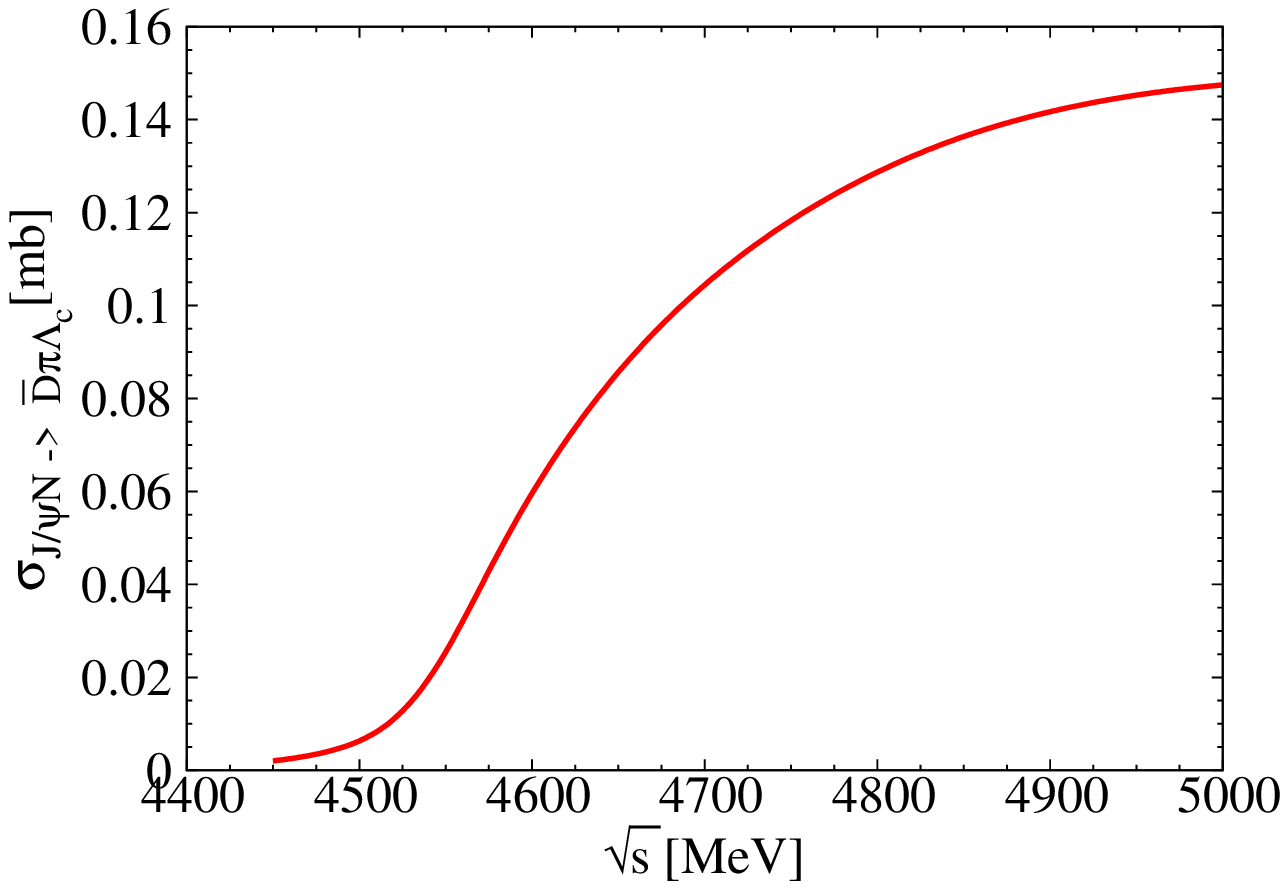}
\includegraphics[scale=0.6]{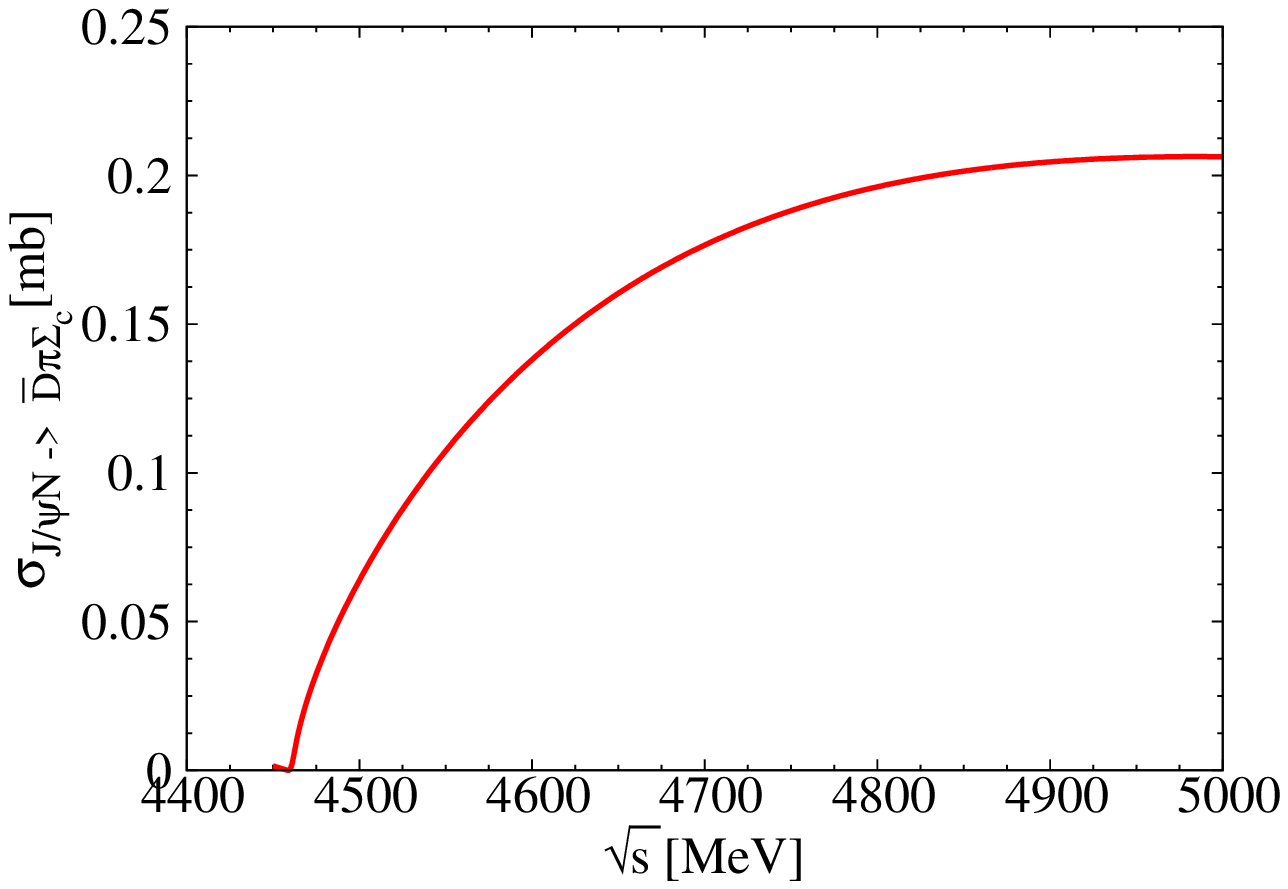}
\caption{The cross section for $J/\psi N \to \bar{D} \pi\Lambda_c (\Sigma_c)$.}\label{figDbpiLS}
\end{figure}

Next we study the reactions $J/\psi N \to \bar{D}\pi\Lambda_c, \bar{D}\pi\Sigma_c$. The diagrams are depicted in Fig. \ref{figFeyn} c).
This process is interesting to study because the $DN$ interaction leads to the $\Lambda_c(2595)$ and $\Sigma_c(2800)$ resonances studied in  \cite{Hofmann:2005sw,Mizutani:2006vq,Tolos:2007vh}, which have the opened decay channels $\pi\Sigma_c$ and $\pi\Lambda_c$ respectively. The scattering matrix for this process is calculated similarly as in the mechanisms of the former section and we find for $J/\psi N \to \bar{D} \pi\Lambda_c$
\be 
\sigma = \frac{M_N M_{\Lambda_c}}{4 p_{J/\psi} s} \int dM_{23} \int_{-1}^1 dcos\theta \frac{p_1 \tilde{p}_2}{(2 \pi)^3} \overline{\sum}\sum |T|^2, \label{sig3body}
\ee
with
\be 
\overline{\sum}\sum |T|^2 = \frac{4}{3} g^2_D\Big[ \frac{(P\cdot p_{\bar{D}})^2}{M_{J/\psi}^2} - m_{\bar{D}}^2 \Big] \Big( \frac{1}{q^2 - m_D^2} \Big)^2 \times \frac{3}{2} \big| T_{DN \to \pi \Lambda_c}^{I=1} \big|^2.
\ee
In Eq. (\ref{sig3body}) $M_{23}$ is the invariant mass of $\pi\Lambda_c$ and $\theta$ the angle between $J/\psi$ and $\bar{D}$, and
\be 
p_1 = \frac{\lambda^{1/2} (s,m_{\bar{D}}^2,M_{23}^2)}{2\sqrt{s}},
\tilde{p}_2 = \frac{\lambda^{1/2} (M_{23}^2,M_{\Lambda_c}^2,m_\pi^2)}{2M_{23}}.
\ee
For the case of $J/\psi N \to \bar{D} \pi\Sigma_c$ we take only the $I=0$ part, which is dominant, and we sum the possible charge processes with this isospin: $J/\psi \;p \to \bar{D}^0 \pi^+\Sigma_c^0$; $J/\psi \;p \to \bar{D}^0 \pi^0\Sigma_c^+$; $J/\psi \;p \to \bar{D}^0 \pi^-\Sigma_c^{++}$. We find
\be 
\overline{\sum}\sum |T|^2 = \frac{4}{3}  g^2_D\Big[ \frac{(P\cdot p_{\bar{D}})^2}{M_{J/\psi}^2} - m_{\bar{D}}^2 \Big] \Big( \frac{1}{q^2 - m_D^2} \Big)^2 \times \frac{1}{2} \big| T_{DN \to \pi \Sigma_c}^{I=0} \big|^2.
\ee
The amplitudes $T_{DN \to \pi \Lambda_c}^{I=1}$ and $T_{DN \to \pi \Sigma_c}^{I=0}$ are evaluated using the model of \cite{Mizutani:2006vq,Tolos:2007vh}. We show the cross section for $\Lambda_c$ and $\Sigma_c$ in the final state in Fig. \ref{figDbpiLS}. The cross sections found are small. The one for $J/\psi N \to \bar{D} \pi\Lambda_c$ is about 30 times smaller than the one for $J/\psi N \to \bar{D} \Lambda_c$, and the one for $J/\psi N \to \bar{D} \pi\Sigma_c$ about five times smaller than that of $J/\psi N \to \bar{D} \Sigma_c$.

\subsection{Transparency ratio}
\label{trans}

We now try to see how we can test this prediction. We can for instance take an electron beam of $10 \gev$ as in the Jefferson lab upgrade and look at
\be 
\gamma A \to J/\psi \; X.
\ee
Depending on what is the elementary production of $J/\psi$, like $\gamma N \to J/\psi N,\, J/\psi \pi N, \cdots$. We will have a range of $J/\psi$ energies in the lab frame which covers the range of energies $4000 \mev - 5340 \mev$. We choose this range because we have the resonance peak for $\sigma_{in}$ in this region. We define the transparency ratio 
\be 
T_A = \frac{\sigma_A(J/\psi)}{A \sigma_N(J/\psi)},
\ee
but it is customary to normalize to a light nucleus like $^{12}C$ and define
\be 
T_A' = \frac{T_A}{T_{^{12}C}}.
\ee
We take several nuclei and evaluate $\sigma_A (J/\psi)$ as a function of $A$. Given the fact that the $J/\psi$ will move in the nucleus essentially forward  in the lab frame of $J/\psi N$, with $N$ a secondary nucleon in the nucleus which we consider at rest, we can use a simple formula derived in \cite{moselz} which gives the transparency ratio as
\be 
T_A = \frac{\pi R^2}{A \sigma_{J/\psi N}} \Big\{ 1 + \big(\frac{\lambda}{R}\big) exp\big[-2 \frac{R}{\lambda}\big] + \frac{1}{2} \big(\frac{\lambda}{R}\big)^2 \big(exp\big[-2 \frac{R}{\lambda}\big] - 1\big) \Big\}. \label{tranew}
\ee
where $\lambda=(\rho_0 \sigma_{J/\psi N})^{-1}$, with $\sigma_{J/\psi N}$ the inelastic cross section of $J/\psi N$. In Eq. (\ref{tranew}) $R$ is the radius of a sphere of uniform density $\rho_0=0.17\; fm^{-3}$ with $R= r_0 A^{1/3}$, $r_0 = 1.143\; fm$ and $A$ the mass number. This formula works remarkably well in comparison with a more accurate one that takes into account the angle dispersion in the laboratory, as we have checked and is also reported in \cite{mariana} in $\eta'$ photoproduction in nuclei.

\begin{figure}
\centering
\includegraphics[scale=0.8]{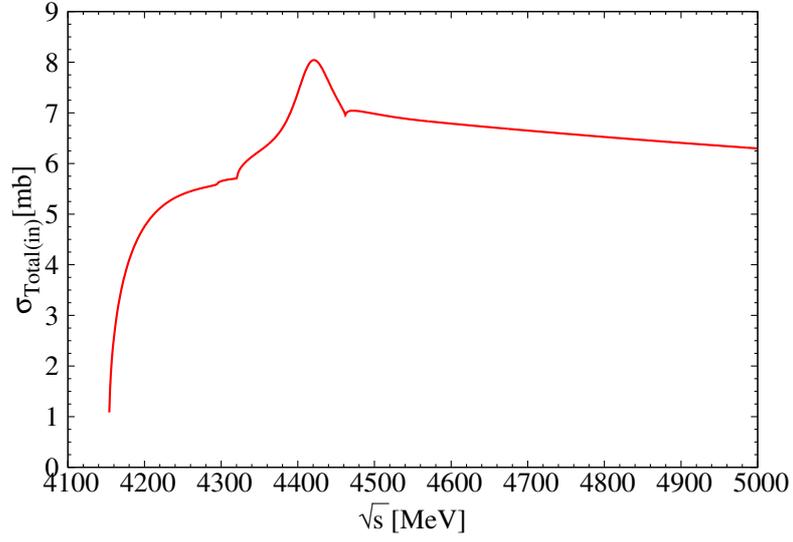}
\caption{The total inelastic cross section of $J/\psi N$.}\label{sigin}
\end{figure}

\begin{figure}
\centering
\includegraphics[scale=0.6]{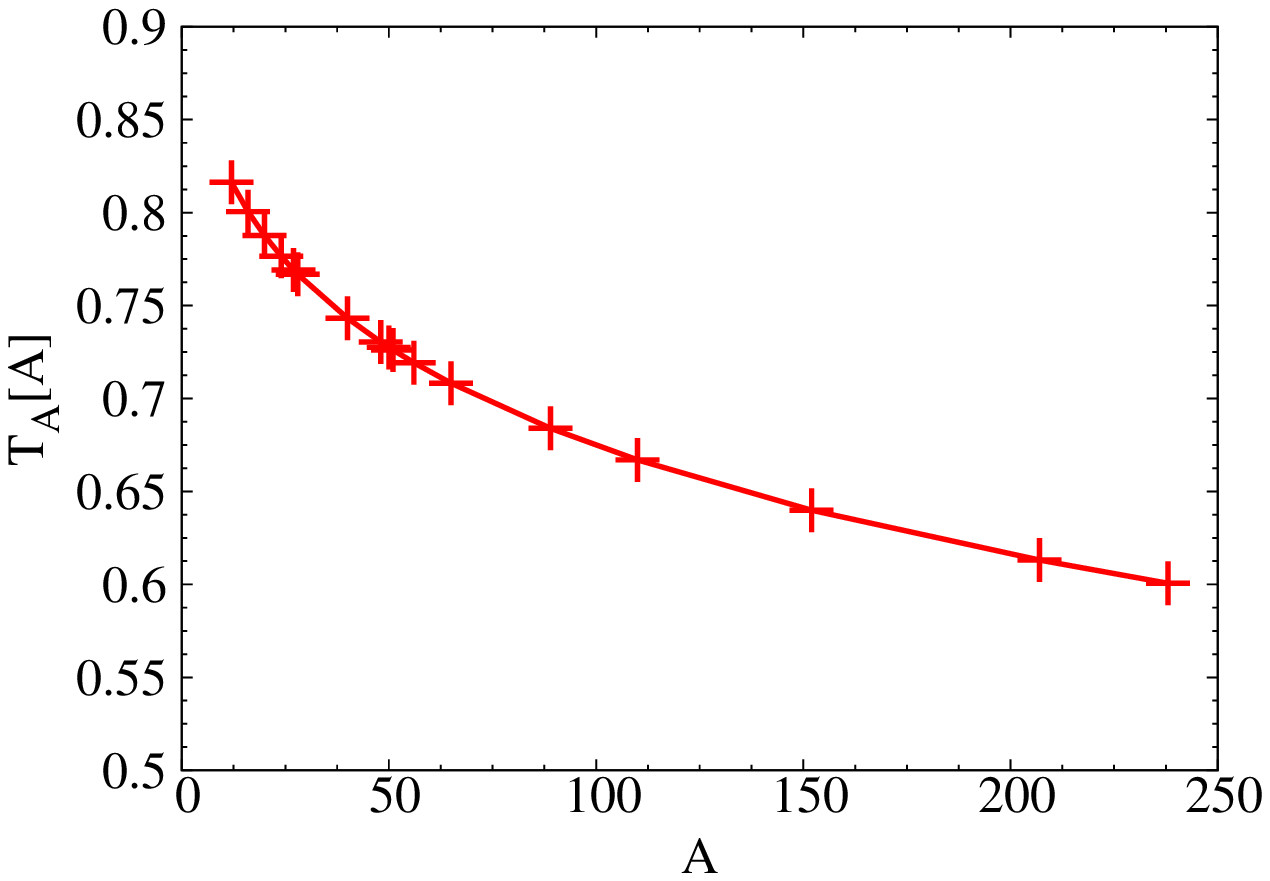}
\includegraphics[scale=0.6]{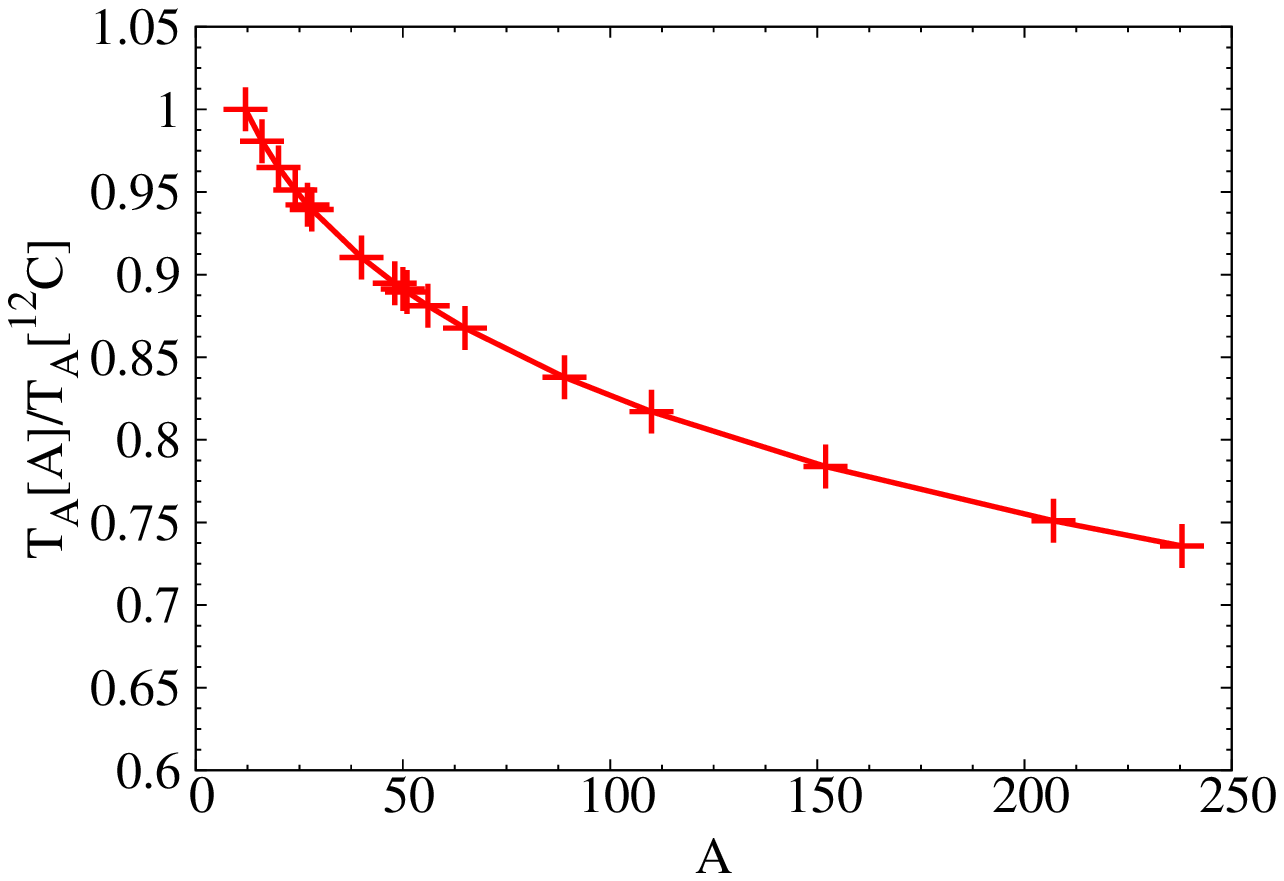}
\caption{The transparency ratio of $J/\psi$ in different nuclei. Left: $T_A$. Right: $T_A/T_{^{12} C}$}\label{figta}
\end{figure}

We plot in Fig. \ref{sigin} the total $J/\psi N$ inelastic cross section, as the sum of all inelastic cross sections from the different sources discussed before.
We can take now various energies of $J/\psi$ and evaluate $T_A$ for this energy as a function of $A$. We do that in Fig. \ref{figta} for $\sqrt{s}=4600\;\mev\;(\sigma_{Total(in)} \simeq 6.8 \;mb)$, a typical energy which is not in the peak of the resonance ($4415$ MeV).
We can see that the values of the transparency ratio are of the order of $0.60-0.70$ for heavy nuclei indicating a depletion of about $30-35$ \%  in $J/\psi$ production in nuclei. Normalized to $T_{^{12}C}$ the ratio go down to $0.75$ for heavy nuclei.

\begin{figure}
\centering
\includegraphics[scale=0.8]{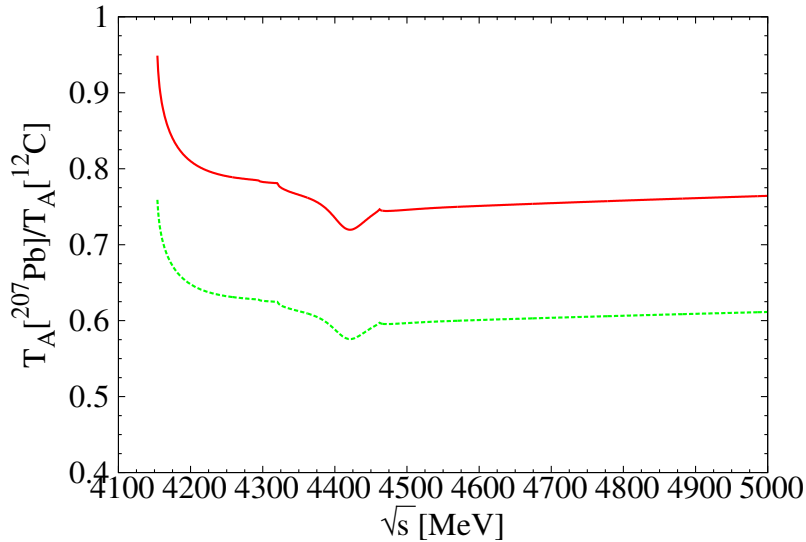}
\caption{The transparency ratio of $J/\psi$ photoproduction  as a function of the energy in the CM of $J/\psi$ with nucleons of the nucleus. Solid line: represents the effects due to $J/\psi$ absorption. Dashed line: includes photon shadowing \cite{Bianchi:1995vb}.}\label{transg}
\end{figure}

In Fig. \ref{transg} we plot the ratio $T_{^{207}Pb} / T_{^{12}C}$ as a function of energy. We can see that the presence of a resonance results into a dip in the ratio of transparency ratios at the energy of the resonance.

It should be noted that the calculation of the transparency ratio done with Eq. (\ref{tranew}) does not consider the shadowing of the photons and assumes they can reach every point without being absorbed. However, for $\gamma$ energies of around $10\gev$, as suggested here, the photon shadowing cannot be ignored. Talking it into account is easy since one can multiply the ratio $T_A'$ by the ratio of $N_{eff}$ for the nucleus of mass $A$ and $^{12}C$. This ratio for $^{208}Pb$ to $^{12}C$ at $E_\gamma =10\gev$ is of the order $0.8$ but with uncertainties \cite{Bianchi:1995vb}. We should then multiply $T_A'(^{208}Pb)$ in Fig. \ref{transg} by this extra factor for a proper comparison with experiment. However this factor does not influence the shape of the results of Fig. \ref{transg} and the dip due to the resonance. 
The small dip in Fig. \ref{transg} would require a high precision experiment to be observed. However, there is one more important reason that makes it not observable, and this is the Fermi motion of the nucleons. Indeed, in the secondary collisions of the $J/\psi$ with nucleons of the nucleus the argument $s$ of the $J/\psi N$ cross section is given by
\be 
s_N=(p_{J/\psi} + p_N)^2 = (E_{J/\psi} + E_N)^2 - (\vec{p}_{J/\psi} + \vec{p}_N)^2,
\ee
while  $E_N \approx M_N$, the term $2 \vec{p}_{J/\psi} \vec{p}_N$ in the expansion of $s$ gives a large span of values of $s$. For this purpose we substitute the $J/\psi N$ inelastic cross section by the one folded over the nucleon momenta
\be 
\sigma (s) \to \bar{\sigma} = \int_{|\vec{p}_N| <p_F} \frac{d^3 \vec{p}_N}{(2 \pi)^3} \sigma (s_N) {\Bigg /} \int_{|\vec{p}_N| <p_F} \frac{d^3 \vec{p}_N}{(2 \pi)^3},
\ee
where $p_F = (3 \pi^2 \rho /2)^{1/3}$ and for $\rho$ we take an average density $\rho \approx \rho_0 /2, \; \rho_0 = 0.17 fm^{-3}$, the nuclear matter density. The differences are minimal if other realistic densities are used.

\begin{figure}
\centering
\includegraphics[scale=0.8]{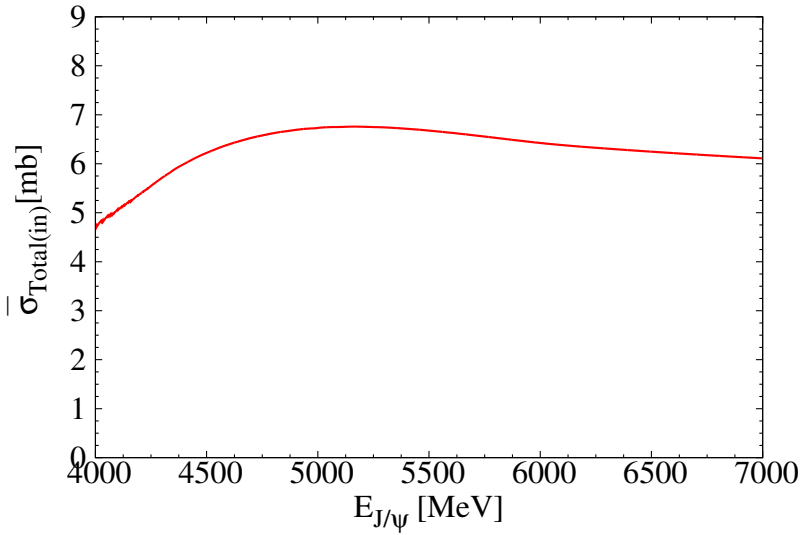}
\caption{The average inelastic cross section of $J/\psi N$.}\label{sigaver}
\end{figure}

\begin{figure}
\centering
\includegraphics[scale=0.8]{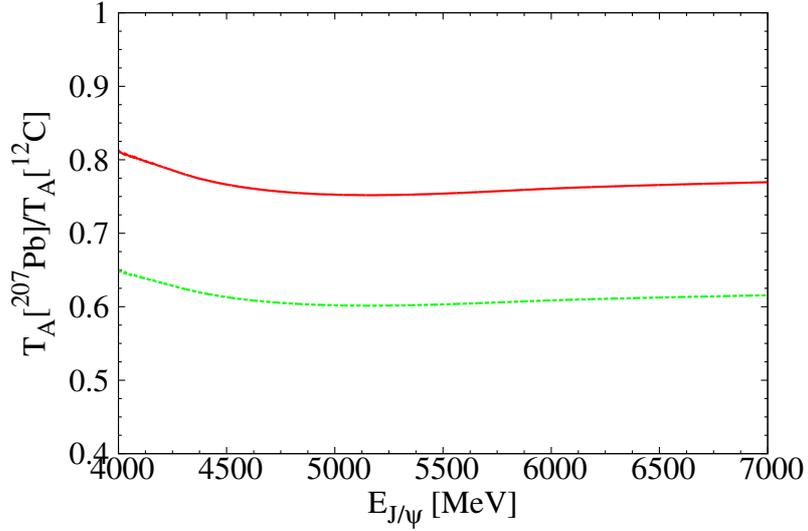}
\caption{The transparency ratio of $J/\psi$ photoproduction  as a function of the energy  of $E_{J/\psi}$ using the averaged $J/\psi N$ cross section over the Fermi sea of Fig. \ref{sigaver}. Solid line: represents the effects due to $J/\psi$ absorption. Dashed line: includes photon shadowing \cite{Bianchi:1995vb}.}\label{transaver}
\end{figure}

The average cross section, $\bar{\sigma}$, is plotted in Fig. \ref{sigaver} as a function of $E_{J/\psi}$. We can see that the peak in Fig. \ref{sigin} is washed away by the effect of Fermi motion. Similarly, we redo the calculations of Fig. \ref{transg} for the transparency ratio using the averaged cross section and we find the results of Fig. \ref{transaver}. There, again, the dip in the transparency ratio has  disappeared, but the values for the $J/\psi$ suppression are essentially the same as before.

As to the values of the transparency ratio for the different nuclei and different energies, even if the suggested experiment studied here has not been done, the values obtained are in line with the rates of suppression found in many experiments \cite{Badier:1983dg,Alde:1990wa,Abreu:1998ee}, where, in spite of using high energies, the $J/\psi$ are produced with momenta in the range studied here.

\section{Conclusions}

 We have investigated different sources of interaction of $J/\psi$ with nucleons in order to obtain the inelastic $J/\psi N$ cross section. First we have used a model recently developed to study the vector-baryon interaction in the charm and hidden charm sectors. This model produces a resonance which couples to $\bar{D}^*\Lambda_c,\bar{D}^*\Sigma_c,J/\psi N$ at $4415$ MeV. The decay of this resonance to light vector-light baryon channels is also incorporated through box Feynman diagrams. Altogheter, it gives contribution to the inelastic part of the $J/\psi N\to J/\psi N$ cross section. We have also considered the transitions $J/\psi N \to \bar{D} \Lambda_c$ or $\bar{D} \Sigma_c$ via $D$-exchange and Kroll Ruderman (contact term) diagrams. These processes give a rate large enough to be observed and dominate for the energies that we consider here ($\sqrt{s}\sim 4100-5000$ MeV). Furthermore, we evaluate the transitions $J/\psi N\to \bar{D}\pi\Lambda_c$ or $\bar{D}\pi\Sigma_c$. However, these latter processes have a small cross section in the range of energies studied here. We find a total inelastic cross section of the order of a few mb, which is sufficient to produce an appreciable suppression of $J/\psi$ in its propagation through nuclei. We then study theoretically the transparency ratio for $J/\psi$ electroproduction in nuclei, for electrons in the range of $10\; \gev$, and find values for the transparency ratio which are in consonance with the typical rates of $J/\psi$ suppression found in most experimental reactions. One interesting side effect is that because of the $J/\psi N$ resonance found theoretically around $\sqrt{s}= 4415\;\mev$, the $J/\psi$ inelastic cross section has a maximum around the energy of this resonance.  
The transparency ratio would have a dip around this energy in principle. However, when the Fermi motion of the nucleus is considered the cross section has to be substituted by its average over the nucleon momenta and the dip is washed away.  The implementation of such an experiment would be rather valuable, providing information on the $J/\psi$ annihilation modes through the nucleonic components of nuclear matter.

\section*{Appendix}
\begin{table}[H]\setlength{\tabcolsep}{0.20cm} \renewcommand{\arraystretch}{1.5}             
\centering
\begin{tabular}{lrrrrrrrr}
\hline\hline
 & $J/\psi N$ & $\bar{D}^* \Lambda_c$ & $\bar{D}^* \Sigma_c$ & $\rho N$ & $\omega N$ & $\phi N$ & $K^* \Lambda$ & $K^* \Sigma$ \\
 $J/\psi N$ & $0$ & $\sqrt{\frac{3}{2}}$ & $-\sqrt{\frac{3}{2}}$ & $0$ & $0$ & $0$ & $0$ & $0$ \\
$\bar{D}^* \Lambda_c$ & & $1$ & $0$ & $-\frac{3}{2}$ & $-\frac{\sqrt{3}}{2}$ & $0$ & $1$ & $0$ \\
$\bar{D}^* \Sigma_c$ & & & $-1$ & $-\frac{1}{2}$ & $\frac{\sqrt{3}}{2}$ & $0$ & $0$ & $1$ \\
$\rho N$ & & & & $-2$ & $0$ & $0$ & $\frac{3}{2}$ & $\frac{1}{2}$ \\
$\omega N$ & & & & & $0$ & $0$ & $\frac{\sqrt{3}}{2}$ & $-\frac{\sqrt{3}}{2}$ \\
$\phi N$ & & & & & & $0$ & $-\sqrt{\frac{3}{2}}$ & $\sqrt{\frac{3}{2}}$ \\
$K^* \Lambda$ & & & & & & & $0$ & $0$ \\
$K^* \Sigma$ & & & & & & & & $-2$ \\
\hline\hline
\end{tabular} \caption{Coefficients $C_{ij}$ in Eq. (\ref{vij}), (\ref{vkl}) and (\ref{eq:box}) for the sector $I=\frac{1}{2},\,S=0$. }\label{cij}
\end{table}

\section*{Acknowledgments}
We would like to thank B. Pires for useful remarks concerning the Fermi motion in the transparency ratio.
This work is partly supported by DGICYT contract number
FIS2011-28853-C02-01, the Generalitat Valenciana in the program Prometeo, 2009/090. We acknowledge the support of the European Community-Research Infrastructure
Integrating Activity
Study of Strongly Interacting Matter (acronym HadronPhysics3, Grant Agreement
n. 283286)
under the Seventh Framework Programme of EU.

\end{document}